\documentstyle[11pt,newpasp,twoside,epsf]{article}
\reversemarginpar

\begin{document}
\title{Warsaw Variability Surveys}
\author{Janusz Kaluzny}
\affil{Copernicus Astronomical Center, ul. Bartycka 18, 00-716 Warsaw\\
\&
Warsaw University Observatory, Al. Ujazdowskie 4, 00-478 Warsaw, Poland}

\begin{abstract}
Four large/medium size photometric surveys are being conducted by
Warsaw astronomers: Warsaw-LCO survey of globular clusters, OGLE,
DIRECT and ASAS. In this contribution we provide short description of
these projects and discuses briefly some results obtained for
pulsating variables.
\end{abstract}

\section{Warsaw-LCO survey of globular clusters}

The main goal of the survey is identification of detached eclipsing
binaries in globular clusters. Such binaries can be used as excellent
distance and age indicators (Paczy\'nski 1997).  Thirteen globular
cluster were surveyed since 1993.  Starting with 1997 most of the data
are collected with the Las Campanas 1.0m Swope telescope equipped with
SITe $2048\times 3150$ CCD giving field of view $14.4\times 22.8$
arcmin$^{2}$. Some additional follow-up observations are conducted on
the 2.5m du Pont telescope.  For a given cluster the total length of
monitoring ranges from 30 to 150 hours with the median value around 70
hours.

Precise and well sampled light curves of RR Lyr and SX Phe stars are
collected as a side result of the survey. These data become available
through Internet (http://sirius.astrouw.edu.pl/\verb"~"jka/)
immediately after results for a given object are published. A list of
monitored clusters and numbers of variables identified in them are
given in Table~1. While monitoring M55 we identified five variables
being likely members of the Sagittarius dwarf galaxy: two RRab stars
and three SX Phe variables.

\begin{center}
\begin{table}
\caption{Pulsating variables detected in the monitored GCs}
\vspace{0.2cm}
\begin{center}
\begin{tabular}{lrrl}
\hline
Cluster  & RR~Lyr & SX Phe & ~~~~~~Reference \\ 
\hline
$\omega$~Cen & 132 & 34     & Kaluzny et al.~(1997, 1997a)\\
NGC~288 & 3 & 5     & Kaluzny et al.~(1997b)\\
NGC~4372 & 0 & 8     & Kaluzny et al.~(1993)\\
NGC~6397 & 1 & 2     & Kaluzny (1997)\\
NGC~6362 & 18 & 4     & Mazur et al.~(1999)\\
         & 35 & 4     & Kaluzny et al.~in prep\\
NGC~6752  & 0 & 3     & Thompson et al.~(1999)\\
M3  & 42 & 1     & Kaluzny et al.~(1998)\\
M4  & 31  & 0      &  Kaluzny et al., in prep.\\
M5  & 99 & 5      & Kaluzny et al.~(1999a, 1999b)\\
    &    &        & Olech et al.~(1999a)\\
M10 & 1 &   3   & Kaluzny et al., in prep.\\
M55 & 13  &           & Olech et al.~(1999)\\
    &     & 24         & Kaluzny et al.~in prep.\\
Ru 106  & 12 & 3     &  Kaluzny et al.~(1995)\\
\hline
Total   & 387    & 96    &  \\
\hline
\hline
\end{tabular}
\end{center}
\end{table}
\end{center}

\subsection{Non radial pulsators in M55}

Our sample of pulsating variables from the globular cluster M55
includes five newly identified RRc stars.  The light curves of three
of these stars exhibit changes in amplitude of over $0.1\;$mag on the
time scale shorter than a week. Detailed analysis indicates that
observed changes are most probably due to non radial pulsations (Olech
et al.~1999).

At least 12 out of 24 SX Phe variables identified in M55 show a
presence of two or more periodicities in their light curves. Table~2
lists principal periods, ratio $P_{1}/P_{2}$ and amplitudes for first
terms of Fourier series calculated for both periods after appropriate
pre-whitening was applied.  The derived values of $P_{1}/P_{2}$
indicate that we are dealing with non radial pulsations. In Fig. 1 we
present light curves obtained on two consecutive night for one of the
multi-periodic SX Phe stars from M55.  We note that also light curve
of an SX~Phe variable identified in M3 by Kaluzny et al.~(1998)
exhibits evidence for non radial pulsations of that star.

\begin{center}
\begin{table}
\caption{Multi-modal SX Phe stars in M55}
\vspace{0.2cm}

\begin{center}
\begin{tabular}{ccccc}
\hline
ID       & P1[d]  &P1/P2& $A_{1}^{1} $ & $A_{1}^{2}$ \\
\hline
 19336 &  0.0388 & 1.017 & 0.020 & 0.009\\
 19407 &  0.0415 & 1.023 & 0.048 & 0.019\\
 25078 &  0.0593 & 1.034 & 0.027 & 0.018\\
 25204 &  0.0370 & 1.026 & 0.014 & 0.006\\
 31780 &  0.0487 & 1.075 & 0.035 & 0.010\\
 34347 &  0.0394 & 1.054 & 0.033 & 0.012\\
 34461 &  0.0438 & 1.023 & 0.025 & 0.011\\
 19084 &  0.0382 & 0.974 & 0.029 & 0.012\\
 12618 &  0.0358 & 1.049 & 0.017 & 0.008\\
 15539 &  0.0370 & 1.067 & 0.014 & 0.007\\
 19176 &  0.0452 & 1.239 & 0.053 & 0.024\\
 19480 &  0.0367 & 1.029 & 0.026 & 0.017\\
\hline
\hline
\end{tabular}
\end{center}
\end{table}
\end{center}

\begin{figure}
\plotfiddle{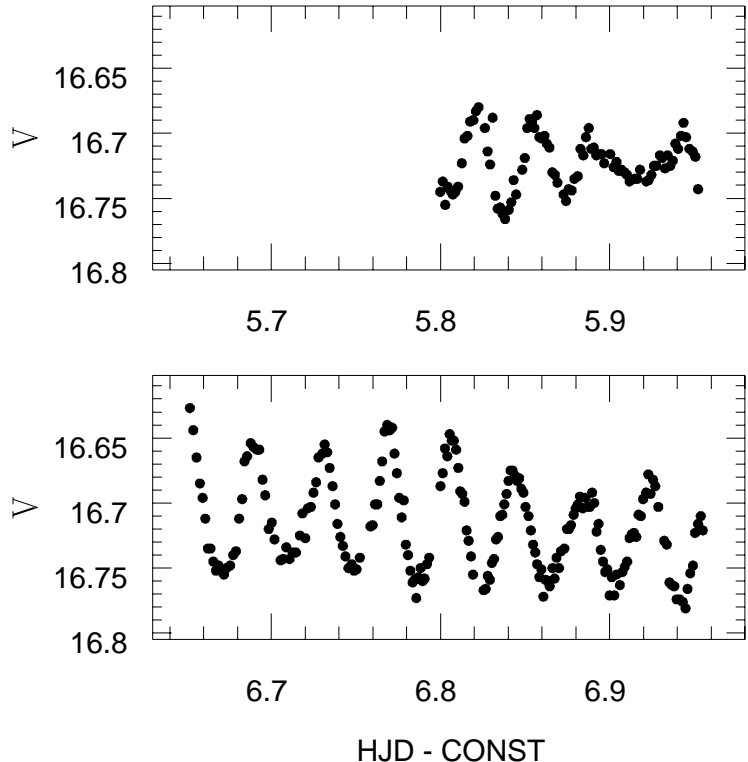}{9.2cm}{0}{85}{85}{-270}{-175}
\caption{Light curve for one of multi-periodic SX Phe stars from M55.}
\end{figure}

\subsection{M3/M55 dichotomy}

Our sample of monitored clusters includes M3 and M55. These objects
have similar metallicities and exhibit similar morphology of their
color-magnitude diagrams (in particular horizontal branches are
similar to each other). Both cluster are rich in blue stragglers yet
show very different relative frequencies of SX Phe stars. This can be
illustrated by comparing numbers of blue stragglers for which we
obtained good quality light curves with numbers of identified SX Phe
stars:

\indent
~~~~~{\bf M55:}  $N_{BS}=40 $, $N_{SX}=24$, $[{\rm Fe/H}]=-1.81$\\
\indent
~~~~~{\bf M3:}~~ $N_{BS}=25 $, $N_{SX}=1$, ~$[{\rm Fe/H}]=-1.57$

Our data for M3 show several blue stragglers located inside
instability strip which do not show any variability exceeding about
0.02 mag in the V band.

\subsection{How complete is  a sample of SX Phe stars identified in GCs?}

Figure~2 shows full amplitude versus period diagram for 96 SX Phe
stars identified by our group in globular clusters.  53 out of these
stars exhibit light curves with full range not exceeding $0.10\;$mag
in the V band. Many variables show amplitudes approaching detection
limit of our survey which, for most clusters reaches
$0.02-0.03\;$mag. That indicates that a significant fraction of SX Phe
stars residing in surveyed cluster was most likely missed.

\begin{figure}
\plotfiddle{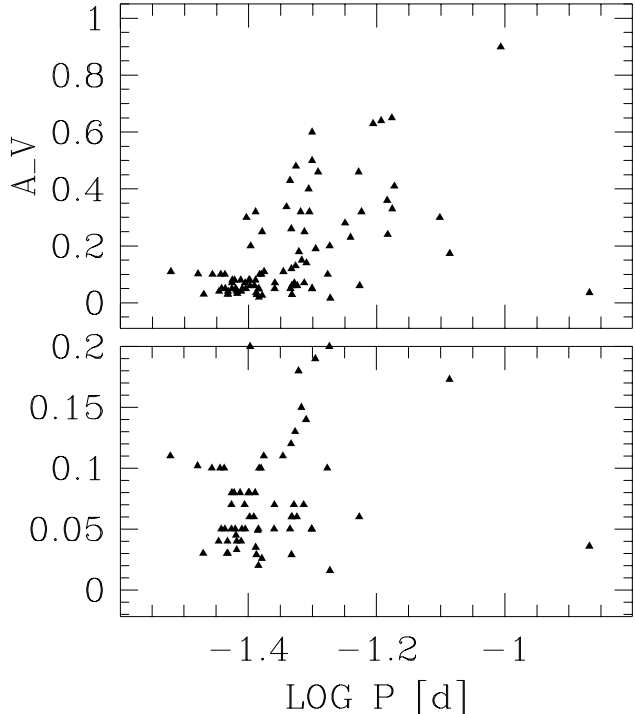}{8.7cm}{0}{58}{58}{-200}{-77}
\caption{Full amplitude versus period for SX Phe stars identified 
during our survey of globular clusters.}
\end{figure}

\section{OGLE-1 survey}

The OGLE project (Udalski et al.~1993;
http://sirius.astrouw.edu.pl/\verb"~"ftp/ogle/) is aimed primarily at
detection and photometry of microlensing events in the Galactic bulge
and LMC/SMC.  The first phase of the survey, refereed here as OGLE-1,
was conducted at Las Campanas Observatory on the 1.0m Swope telescope
during 4 seasons covering the period 1992-1995.  About $2\times
10^{6}$ stars in 20 $15\times 15$ arcmin$^{2}$ fields toward the
Galactic bulge were monitored. 20 microlensing events were detected
(Wozniak \& Szymanski 1998). Five parts of the Variable Star Catalog
were published, containing 2861 stars from the Galactic bulge (Udalski
et al.~1997a). That catalog includes 269 pulsating objects, mostly RR
Lyr stars.

Several side projects were attempted by the OGLE-1 team.  We list here
these of them which yield some possibly interesting results for
pulsating variables. Mateo et al.~(1995) searched with success for RR
Lyr stars belonging to the Sagittarius dwarf galaxy. They reported VI
photometry for 7 variables from that galaxy.  V band data for 226 RR
Lyr stars from the Sculptor dSph galaxy were obtained by Kaluzny et
al.~(1995a).  That galaxy may prove to be an ideal target for
calibration of the luminosity-metallicity relation for RR Lyr stars.
Population of stars hosted by Sculptor shows significant range of
metallicities and the interstellar reddening toward the galaxy is very
low. Kaluzny et al.~(1996, 1997) identified 34 SX Phe stars in the
globular cluster $\omega $~Cen. V band data for 141 RR Lyr stars (33
newly identified) and Pop II Cepheids in the same cluster were
published by Kaluzny et al.~(1997a).
  
In 1996 the OGLE project entered its second phase known as OGLE-2
(Udalski, Szymanski \& Kubiak 1997).  OGLE-2 results are described in
this volume by Paczy\'nski (1999).

\section{DIRECT}

The DIRECT project (http://cfa-harvard.edu/\verb"~"kstanek/DIRECT/)
aims at determination of distance to M31 and M33 galaxies by using
detached eclipsing binaries and Cepheids.  The project is currently
conducted by a group including astronomers from CfA and Warsaw.  About
200 nights on the 1.2m FLWO and 1.3 MDM telescopes were used between
September 1996 and November 1999 to search both galaxies for variables
suitable for more detailed follow-up.  So far five catalogs of
variables in M31 were released (Kaluzny et al.~1998a, 1999; Stanek et
al.~1998, 1999; Mochejska et al.~1999).  410 variables (most of them
new) were identified, including 206 Cepheids and 48 eclipsing
binaries.  Photometry of many RV Tau and LPV stars was also
reported. The remaining catalogs shall be released over the coming
year. Some results of DIRECT concerning Cepheids are presented in this
volume by Sasselov (1999)

\section{ASAS}
ASAS (the All Sky Automated Survey; Pojmanski 1997;\\
http://www.astrouw.edu.pl/\verb"~"gp/asas) is a project 
which ultimate goal is low cost monitoring of the whole 
sky on a nightly basis down to about 15 magnitude.
A prototype robotic telescope consisting of the 135mm telephoto lens,
off-the-shelf CCD camera  (512 x 768 pixels)
and small automated mount was set
up at the Las Campanas Observatory in April 1997. 
It has been monitoring 24 selected fields covering about 150 
deg$^2$ of the sky.
Usefull photometry in the I band was obtained for over 45000
stars brighter than 13 magnitude. The first two month of observation
revealed 126 short period variables (Pojma{\'n}ski 1998), 
of which 70 \%  were previously unknown. The catalogue includes 
several newly identified RR Lyr stars and Cepheids. 
In 1998 the survey was extended to cover additional 150 deg$^2$ and 
the updated version of the catalogue will include data for a large 
number of pulsating stars with periods  up to 300 days
(Pojmanski 1999, private communication).
At the end of 1999 two new robotic telescopes with 2K$^2$ CCD's
will be installed at Las Campanas. 
\acknowledgments
JK was supported by the Polish Committee of Scientific Research through
grant 2P03D000.00 and by NSF grants AST-9528096 and AST-9819787
to Bohdan Paczy\'nski.

\end{document}